\documentclass{sig-alternate}
\usepackage{mathptmx} % This is Times font

\usepackage{fancyhdr}
\usepackage[normalem]{ulem}
\usepackage[hyphens]{url}
\usepackage[sort,nocompress]{cite}
\usepackage[final]{microtype}
\usepackage[keeplastbox]{flushend}

\usepackage{amsmath,amssymb,amsfonts}
\usepackage{algorithmic}
\usepackage{graphicx}
\usepackage{textcomp}
\usepackage{xcolor}
\usepackage{multirow}
\usepackage{enumerate, pifont, todonotes}
\usepackage{array}
\usepackage{threeparttable}
\usepackage{caption}
\usepackage{listings}
\usepackage{adjustbox}
\usepackage{colortbl}
\definecolor{tablegray}{gray}{.95}
\definecolor{mygreen}{rgb}{0,0.6,0}
\definecolor{mygray}{rgb}{0.9,0.5,0.5}
\definecolor{mymauve}{rgb}{0.58,0,0.82}

\DeclareCaptionType{codetype}[Listing][List of mytype]

\usepackage[T1]{fontenc}
\usepackage[utf8]{inputenc}
\usepackage{authblk}
\usepackage[bookmarks=true,breaklinks=true,letterpaper=true,colorlinks,linkcolor=black,citecolor=blue,urlcolor=black]{hyperref}

\newcommand{\red}[1]{\textcolor{red}{#1}}

\title{A Lightweight Isolation Mechanism for Secure Branch Predictors} 
\begin{document}

% \vspace{-2.0cm}
\author{
% \IEEEauthorblockN{
Lutan Zhao\textsuperscript{\dag}, 
Peinan Li\textsuperscript{\dag},
Rui Hou\textsuperscript{\dag}\thanks{Corresponding author: Rui Hou (hourui@iie.ac.cn)} , 
Michael C. Huang\textsuperscript{\ddag},
Jiazhen Li\textsuperscript{\dag},
Lixin Zhang\textsuperscript{\S},
Xuehai Qian\textsuperscript{\P} and
Dan Meng\textsuperscript{\dag} \\
% } \\
% \IEEEauthorblockA{
\textsuperscript{\dag}State Key Laboratory of Information Security, Institute of Information Engineering, CAS\\
and University of Chinese Academy of Sciences.
% }
% \IEEEauthorblockA{
\textsuperscript{\ddag}University of Rochester.\\
% }\\
% \IEEEauthorblockA{
\textsuperscript{\S}Institute of Computing Technology, CAS.
% }
% \IEEEauthorblockA{
\textsuperscript{\P}University of Southern California.
% }
}

\maketitle

\pagestyle{plain}

% \maketitle
% \thispagestyle{firstpage}
% \pagestyle{plain}

\newcommand\note[1]{\textcolor{blue}{\{\textbf{Note:} {\em#1}\}}}
\let\oldtodo\todo
\newcommand\disc[1]{\oldtodo[inline, color=green]{#1}}
\newcommand{\fixme}[1]{\red{{\Large FIXME:} {\bf #1}}}
\renewcommand{\todo}[1]{\oldtodo[inline]{#1}}
\newcommand\zhaolutan[1]{\textcolor{green}{\{\textbf{zhaolutan:} {\em#1}\}}}
\newcommand\lipeinan[1]{\textcolor{purple}{\{\textbf{lipeinan:} {\em#1}\}}}

\newcommand{\donotshow}[1]{}

%%%%%% -- PAPER CONTENT STARTS-- %%%%%%%%
\begin{abstract}
Recently exposed vulnerabilities reveal the necessity to improve the security of branch predictors. Branch predictors record history about the execution of different programs, and such information from different processes are stored in the same structure and thus accessible to each other. This leaves the attackers with the opportunities for malicious training and malicious perception. Instead of flush-based or physical isolation of hardware resources, we want to achieve isolation of the \textbf{content} in these hardware tables with some lightweight processing using randomization as follows. (1) \textbf{Content encoding}. We propose to use hardware-based thread-private random numbers to encode the contents of the branch predictor tables (both direction and destination histories) which we call XOR-BP. Specifically, the data is encoded by XOR operation with the key before written in the table and decoded after read from the table. Such a mechanism obfuscates the information adding difficulties to cross-process or cross-privilege level analysis and perception. It achieves a similar effect of logical isolation but adds little in terms of space or time overheads. (2) \textbf{Index encoding}. We propose a randomized index mechanism of the branch predictor (Noisy-XOR-BP). Similar to the XOR-BP, another thread-private random number is used together with the branch instruction address as the input to compute the index of the branch predictor. This randomized indexing mechanism disrupts the correspondence between the branch instruction address and the branch predictor entry, thus increases the noise for malicious perception attacks. Our analyses using an FPGA-based RISC-V processor prototype and additional auxiliary simulations suggest that the proposed mechanisms incur a very small performance cost while providing strong protection.

\end{abstract}

%%%%%%%%%%%%%%%%%%%%%%%%%%%%%%%%%%%%%%%%%%%%%%%%%%%%%%%%%%%%%%%%%%
\section{Introduction}

In modern processors, branch prediction is crucial in effectively exploiting parallelism of sequential programs for high-performance execution. However, recently exposed vulnerabilities reveal the necessity to improve the security of branch predictors in mainstream commercial processors~\cite{kocher2018spectre,evtyushkin2018branchscope,evtyushkin2015covert,aciiccmez2007power}. Take Spectre V2 vulnerability~\cite{kocher2018spectre} as an example: Through purposeful training of the branch predictors, an adversary can change the control flow of a victim to incorrect speculative paths which in turn reveals unauthorized data. Another example is the BranchScope attack~\cite{evtyushkin2018branchscope}, where history information in the branch predictor can be perceived by the attacker to infer sensitive data after the victim has executed. The root cause of these vulnerabilities is that modern processors generally adopt the design principle of resource sharing, and branch predictor is one of the typical examples. From a security perspective, resource sharing means a possible attack surface. Branch predictors record history about the execution of different programs, and such information from different processes is stored in the same structure and thus accessible to each other. This leaves the attackers with the opportunities for malicious training and malicious perception.

One kind of defensive idea is to avoid letting key instructions leave traces in the branch predictors. For example, one countermeasure~\cite{agosta2007countermeasures} is to convert secret-dependent branch instructions into indirect jumps or other computation instructions to prevent the leakage of sensitive information. As another example, Intel processors allow software to set IA32\_SPEC\_CTRL.IBRS to 1 after 
a transition to a more privileged predictor mode~\cite{Intel2018specmitigations}.  Predicted targets of indirect branches executed in that predictor mode cannot be set by execution in a less privileged predictor mode. However, it is difficult to guarantee full coverage of sensitive branches through manual or automated analyses. 

In contrast, isolation is a fundamental way to improve the security of branch
predictors. Existing proposals can be largely separated into two categories.
\ding{172}
Logical isolation aims at preventing attacks on the shared hardware resource.
Attaching the thread ID to each entry can help eliminate malicious reuse
across threads. But it cannot avoid contention. Another mechanism is to
flush the whole history table each time a context switch or privilege
switch (e.g., system call). Evtyushkin et al. proposed to do so by software
method~\cite{evtyushkin2016understanding}. Clearly, this approach introduces
a large context switching cost (e.g., 1.2ms per switch in the experiment).
Naturally, hard-wiring flush support can bring the cost of flushing down,
which was simulated by Sangho et al.~\cite{lee2017inferring}. Even with
lowered costs to perform the flush operation, decrease of the performance
benefits of branch predictors is yet another cost, which becomes significant
in SMT architectures.
\ding{173} Physical isolation can be implemented by allocating separate branch
tables for different threads or privilege levels. BRB~\cite{vougioukas2019brb}
is a state-of-art hardware implementation that provides individual history
tables for different programs. Although BRB tries to limit hardware cost,
it is in general impractical to assign separate tables to all threads and
privilege level combinations. Also, physical isolation alone is insufficient
as storage is still multiplexed in time by different threads.

In short, existing methods are not satisfactory. We thus set out to seek a
more lightweight solution that achieves isolation of the content with similar
or better qualities than these prior approaches.
To the best of our knowledge, this paper is the first to propose the XOR-based lightweight isolation mechanism for branch predictors, which can effectively prevent branch-predictor induced malicious code execution and prevent spying on the history information of branch predictors. Our approach has the advantage of simple implementations and better suitability for SMT architecture than predictor flushing upon context switch. It is also more cost effective than private predictors or hardware backups. Overall, this paper makes the following contributions:

\begin{itemize}
\item We propose a lightweight XOR-based Isolation mechanism (\emph{XOR-BP}) for branch predictors. Upon context switch or privilege changes, the hardware dynamically generates a new thread-private random number. When updating the branch predictor (direction and/or address), the predictor's update is XORed with the thread-private random number before saved to the table. Similarly, when reading the predictor table, the current random number will be used again to decode the stored result. By changing the thread-private random number periodically (upon context switch or privilege changes), this lightweight mechanism effectively achieves content-level isolation among multiple threads.

\item We propose a randomized index mechanism of the branch predictor (\emph{Noisy-XOR-BP}). Based on the \emph{XOR-BP}, another thread-private random number is used together with the branch instruction address as the input to compute the index of the branch predictor. This randomized indexing mechanism disrupts the correspondence between the branch instruction address and the branch predictor entry, thus increases the noise for malicious perceived attacks.

\item We present a detailed evaluation for our proposed designs. For single-threaded core, we implemented the above isolation mechanisms on an FPGA prototyping system of a RISC-V out-of-order processor for a realistic evaluation of the performance impact. For SMT microarchitectures, we modeled and evaluated the isolation mechanisms on a Gem5-based simulator, and we investigate its performance impacts for different branch predictors, including the latest TAGE\_SC\_L~\cite{seznec_TAGE_SC_L} predictors.

\end{itemize}

In the following, we analyze existing attacks on branch predictors (Section ~\ref{vulnerabilities});
discuss the threat model (Section ~\ref{threat}); introduce the defense strategy (Section ~\ref{Strategy}) and the design details of the countermeasure (Section ~\ref{XOR}); analyze the performance impacts (Section ~\ref{evaluation}); summarize related work (Section ~\ref{related}); and conclude this paper (Section ~\ref{conclusion});

% % that's all folks

\section{Understanding Attacks Via Branch Predictors}
\label{vulnerabilities}
\subsection{Two types of attacks due to resource sharing}

Conventional branch predictor design allows different processes to use the
same hardware resources for branch prediction. This creates side channel
just like those exploited in a cache based side channel attack. it allows
the attacker to prime the predictor in a certain fashion to facilitate the
revelation of a victim's sensitive information. Additionally, the attacker can
achieve malicious training in order to influence the victim's (speculative)
execution, which in turn enables or enhances the victim's information leak. A
number of different types of attacks have been constructed:

\textbf{(1) Reuse based attacks}: In structures such as PHT (Pattern History Table), different programs directly access the common resource. Entries set by one process influence another. This type of attack is therefore analogous to reuse based cache attacks~\cite{liu2014random}. There are three typical examples.

\ding{172} BranchScope attack~\cite{evtyushkin2018branchscope}: An attacker
first locates the shared PHT entry of the secret-dependent branch of the
victim and sets its saturating counter to a specific state, such as \emph{Weak
Taken}. The contents of the branch predictor are updated after the victim's
target branch instruction is executed. After switching back to the attacker's
program, the victim's update to the branch predictor manifests as a measurable
difference in execution time. The attacker can thus sense the direction of
the target instruction and infer the victim's execution path.

\ding{173} Pure malicious training like Spectre V1 and V2~\cite{kocher2018spectre}: Instead of exploiting the side channel of predictor to obtain leaked information, an attacker carefully trains the shared predictor, either PHT or BTB, to induce victim thread into incorrect speculative paths which in turn reveals unauthorized data via cache side channel.

%\item 
\ding{174} Branch Shadowing attack~\cite{lee2017inferring} makes a shadow of victim code and measures whether target address in BTB left by victim branch accelerates the execution of shadow branch, by which the attacker can observe the direction of victim branch in SGX.
%\end{enumerate}

\textbf{(2) Contention based attacks}: In structures more similar to a cache, such as the BTB, an attack similar to contention based cache attack can be mounted. 
One condition for constructing such attacks is that contention result in eviction of the old record.
An attacker can learn about the execution of the target branch instructions of 
a victim by sensing whether contention happens or not for the corresponding entry of branch prediction table.
Because different branches in typical PHT use and update the same table, rather than evicting others' history, there is no contention based attacks.

%\begin{enumerate}
%\item 
In an SBPA attack~\cite{aciiccmez2007power,aciiccmez2007predicting}, the
attacker first occupies all the entries in the same set (multiple ways) in
the BTB corresponding to the victim's target branch instruction. When the
victim executes, the target branch will have a BTB miss and thus predicted as
\emph{Not Taken}. According to the BTB's update mechanism, the BTB will be
updated if and only if the target branch is \emph{Taken}. An update of the BTB
will evict an entry primed by the attacker, resulting in measurable difference
of execution time, thus allowing the attacker to learn the execution results
of the target branch of the victim.
    %\item 
    
    An attacker can traverse its own address space and measuring execution time to
    infer the eviction of its branch from the BTB. In the case where only 
    one branch is routinely evicted, then the victim must have a branch with the same index
    and (partial) tag. This allows the attacker to
    infer virtual addresses of other threads and break KASLR~\cite{evtyushkin2016jump}.

In summary, the fundamental source of these attacks is that current branch predictors do not have thorough isolation between different processes and privileges.

\subsection{Common steps for attacks on branch predictor}

By analyzing these two types of attacks, the following common steps can be observed during a typical attack:

\textbf{Step 1: Locate phase}. An attacker first needs to locate the entry of
branch predictor for the victim's target branch instruction. Given the
rather fixed indexing design of typical predictor implementation, this is
relatively straightforward.

\textbf{Step 2: Prime phase}. Next, the attacker needs to prepare state for
the target entry. This includes priming the whole set in order to sense
eviction; or set a particular value in order to achieve either malicious
training, or to permit future observation of the content changes due to the
execution of the target branch.

\textbf{Step 3: Probe phase (optional)}. After the target branch is executed,
if the attacker is to obtain information about the target branch, it needs to
probe the status of the target entry, usually via execution time analysis.

\section{Threat Model}
\label{threat}

This paper has the following assumptions: The attacker thread and the victim thread can run on the same processor core. An attacker can know the source code and address layout of the victim. An attacker has the ability to run the victim program in single-step mode~\cite{evtyushkin2018branchscope}.

This paper focuses on the defense against the reuse based attacks and the contention based attacks, which cause malicious training, priming, and perception across different processes and privileges. In addition, branch predictors may also have side-channel leakage in the event of a mis-prediction. It should be noted, however, misprediction 
and thus mis-speculation are perhaps inevitable.
This paper does not consider the defense against all speculative execution related vulnerabilities.

A typical branch predictor consists of a set of  history/pattern tables, which work together to predict the direction of a conditional branch (Pattern History Table, or PHT), the addresses of indirect branches (Branch Target Buffer, or BTB), and the addresses of function return instructions (Return Address Stack, or RAS). Many commercial processors have adopted a thread-private RAS structure for SMT core, but they still use a shared design for PHT and BTB~\cite{Intel2018manual}. Therefore, this paper mainly studies how to isolate PHT and BTB. Nevertheless, our proposed methods still apply to shared RAS.

% that's all folks

\section{Defense Strategy}
\label{Strategy}
This paper focuses on securing branch predictors against the reuse based and contention based attacks. As can be seen from the analysis in Section~\ref{vulnerabilities}, these attacks primarily exploit the design vulnerability that branch predictors allow concurrent and fine-grain shared accesses from multiple threads and multiple privilege spaces. Our approach is to provide sufficient isolation in the predictor between threads or privilege levels. We first examine existing isolation proposals.

\subsection{Analysis of existing isolation mechanisms}

{\bf (1) Logical isolation} aims at preventing attacks on physically shared branch predictor and only allows its owner to access.
This could take the form of adding some sort of thread ID to each entry.
For example, a PHT of a commercial processor typically has around 4K entries and each being a 2-bit counter.
Adding ASID information (12bit in an Intel processor~\cite{Intel2018manual}) to every entry is a costly approach.
However, there still exists contention based attacks if the attacker can discern whether its history has been evicted by victim branch.
Flushing the predictor completely upon context switches or privilege changes is an alternative logical isolation mechanism, which is named as \emph{Complete Flush} in this paper.
We investigate its performance impacts on our evaluation platform, which is described in details in Section~\ref{evaluation}.

\begin{figure}[ht]
% \vspace{-0.8em}
% \setlength{\abovecaptionskip}{-0.2em}
% \setlength{\belowcaptionskip}{2ex}
\centering
  \includegraphics[width=0.48\textwidth]{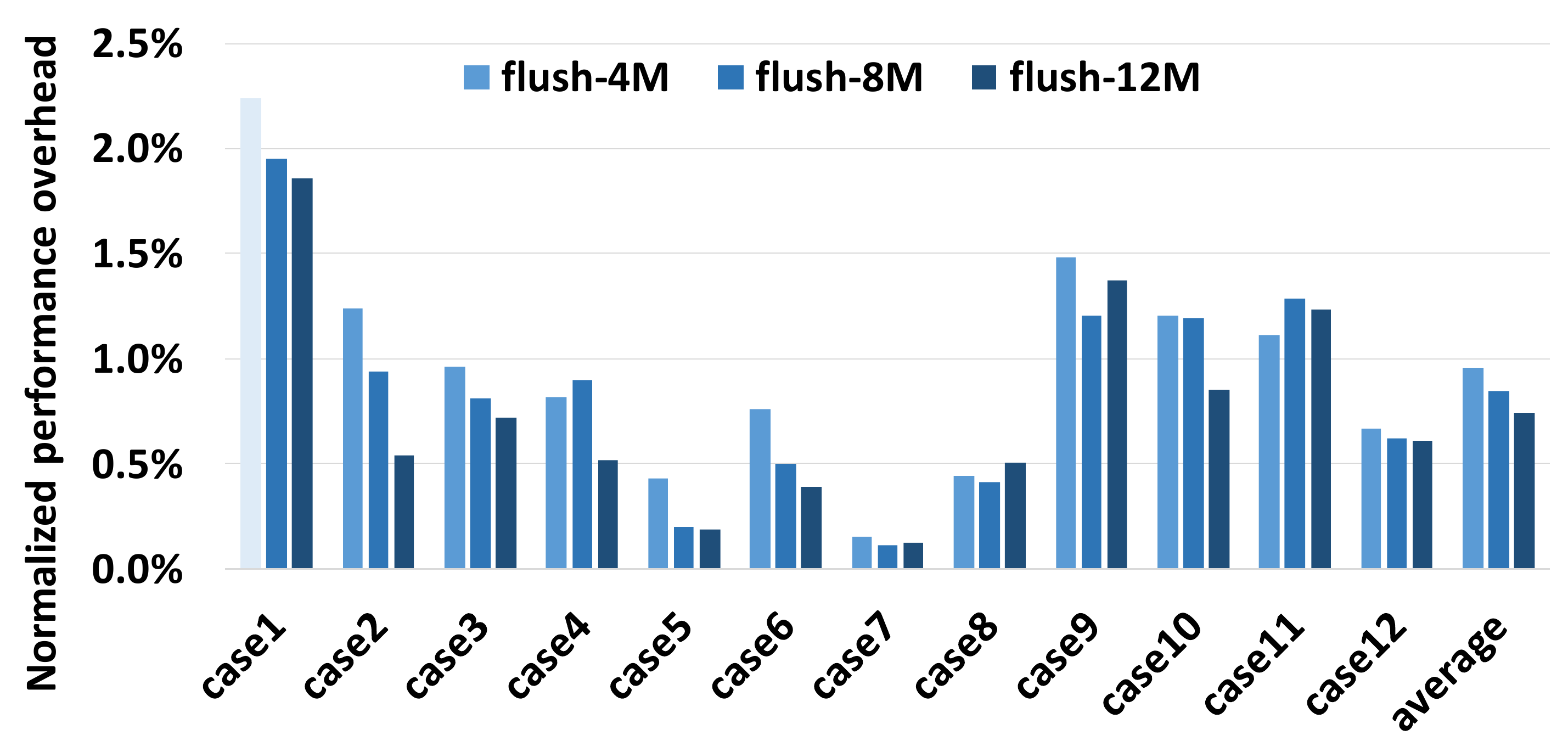}
\caption{Performance overhead of flushing branch predictor on single-threaded processor. \footnotesize{(\emph{flush-4M} means the predictor is flushed every 4 million cycles)}. }
\label{flush_single}
\vspace{-0.8em}
\end{figure}

There are three interesting observations:

\textbf{Observation 1: The performance impact of flush methods on single-threaded core is insignificant.}
For a 2GHz single-threaded core, when the context switching frequency is 250Hz (typical context-switching frequency in Linux), the average performance loss of flushing predictors upon switching is less than 1\% in comparison with the baseline (without isolation) as shown in Figure ~\ref{flush_single}.
It indicates that the program usually has enough time in each scheduled execution window to amortize the warm-up of branch predictor, which is consistent with other findings~\cite{lee2017inferring}.

\begin{figure}[ht]
% \vspace{-0.8em}
% \setlength{\abovecaptionskip}{-0.2em}
% \setlength{\belowcaptionskip}{-5em}
\centering
  \includegraphics[width=0.48\textwidth]{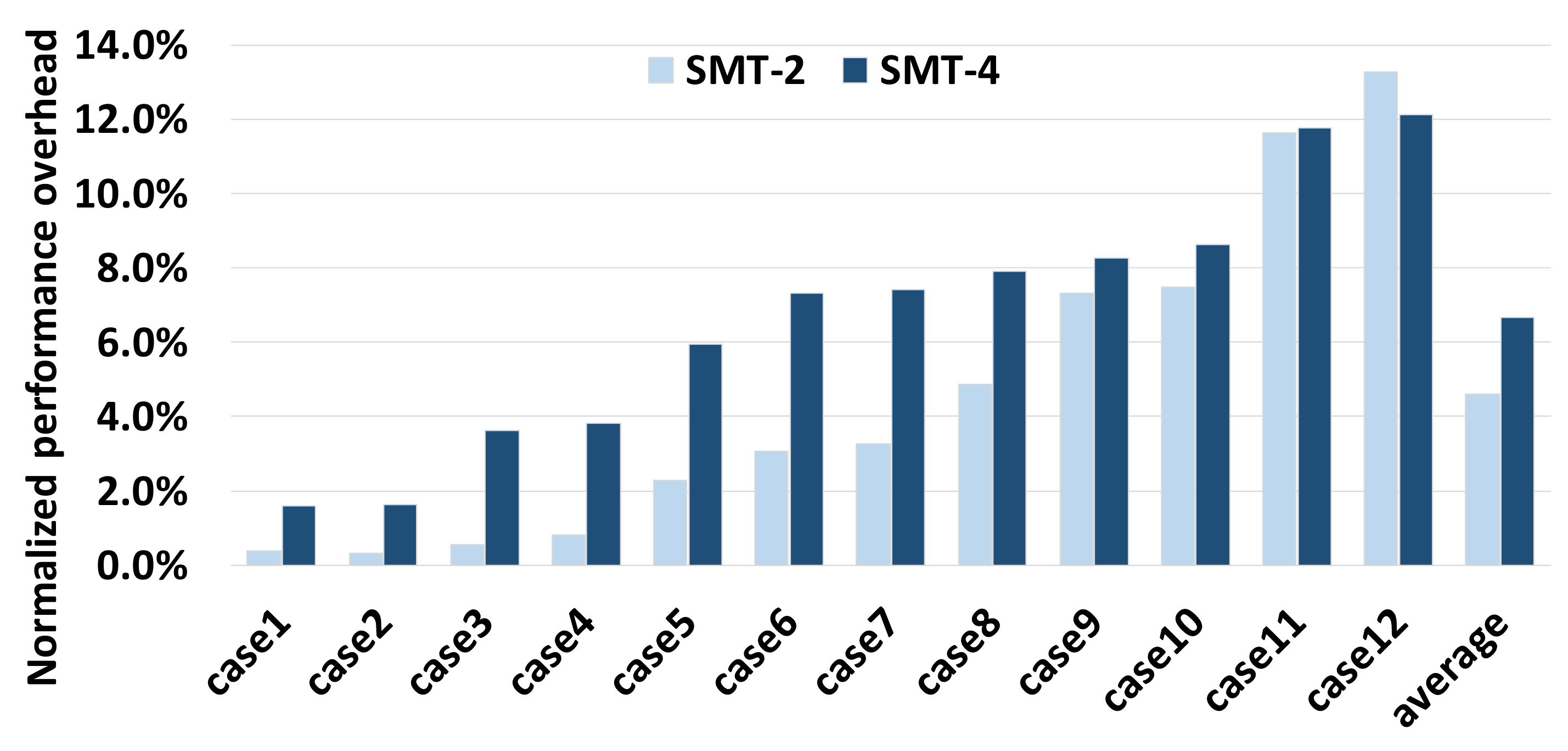}
\caption{Performance overhead of flushing branch history on an SMT core.}
\label{flush_SMT}
\vspace{-0.8em}
\end{figure}

\textbf{Observation 2: The performance loss of \emph{Complete Flush} on an SMT core gets worse since flushing from different threads interfere with each other.
Furthermore, \emph{Complete Flush} cannot prevent attacks on SMT cores.}
Figure ~\ref{flush_SMT} shows a significant increase in performance loss to flush in
an SMT core compared to a single-threaded core. 
Increasing the number of threads causes more performance degradation.
And it cannot provide sufficient isolation to prevent sharing and preemption of history from other hardware threads.

\begin{figure}[ht]
% \vspace{-0.8em}
% \setlength{\abovecaptionskip}{-0.2em}
% \setlength{\belowcaptionskip}{2ex}
\centering
  \includegraphics[width=0.48\textwidth]{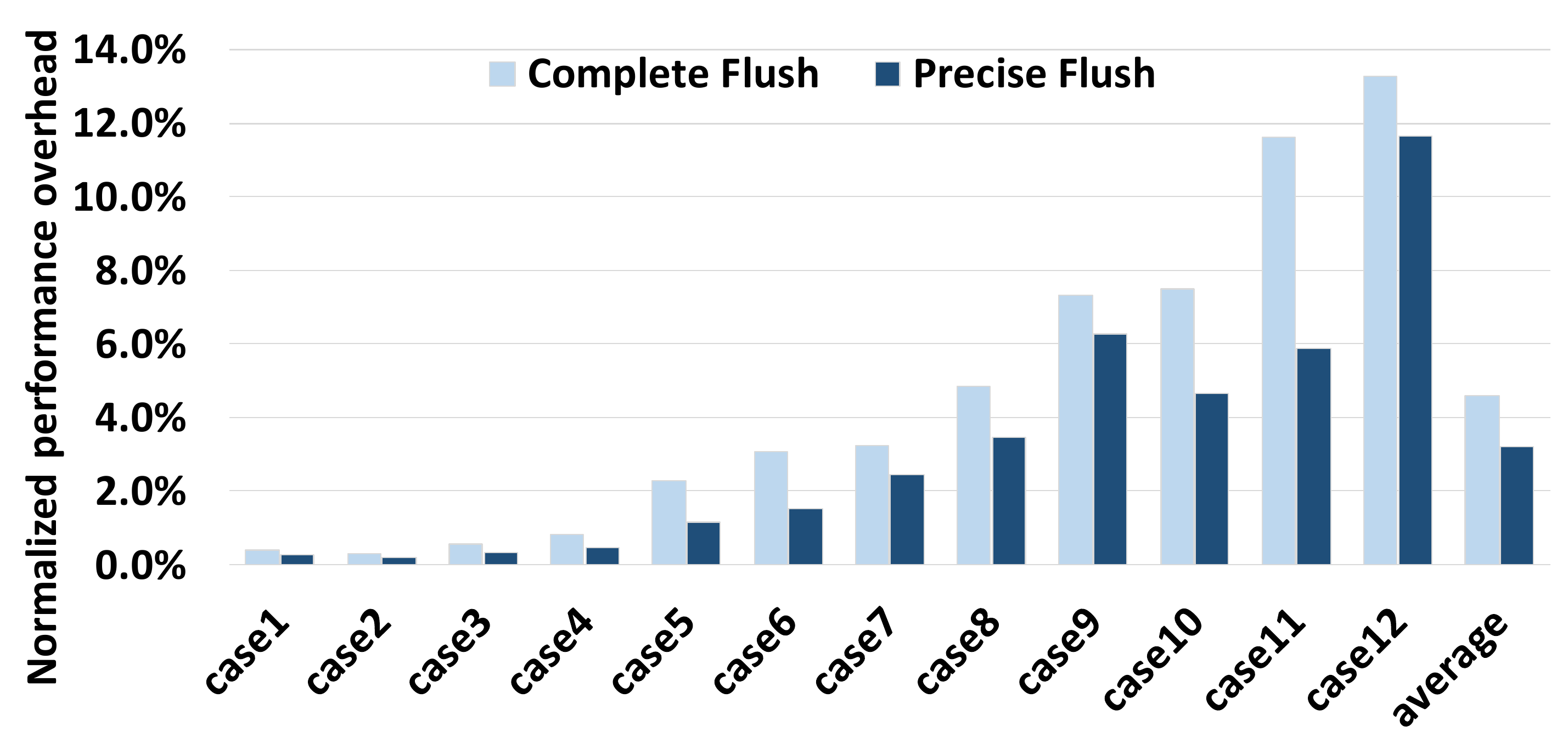}
\caption{Comparison between \emph{Complete Flush} and \emph{Precise Flush} in SMT-2 (Normalized to baseline without any mechanism).}
\label{SMT_flush_accurate}
\vspace{-0.8em}
\end{figure}

\textbf{Observation 3: At the cost of complex hardware implementations, a more precise flush mechanism is better but not sufficient.
And this method may introduce new security issue for PHT.}
It is tempting to think that if we can replace a whole-structure flush with a more precise
flush mechanism, the performance degradation problem will be solved. Unfortunately, this
does not appear to be the case. We have simulated a design in which each entry of branch
prediction resource is augmented with thread ID properly managed to ensure only
entries used by a particular thread are flushed when the thread is swapped out.
We show performance comparison of such a more precise flush mechanism (dubbed
\emph{Precise Flush}) vs the more basic variant (\emph{Complete Flush})
in Figure~\ref{SMT_flush_accurate}. We see that the performance loss 
does reduce but remains elevated. Considering the extra storage space and complexity
involved to carry out the more precise flush, it is hard to see this as a 
satisfactory solution, especially for a 2-bit pattern history table.
Furthermore, such a flush mechanism still cannot protect against 
a contention based attack.

{\bf (2) Physical isolation} means allocating separate branch tables for different threads and different privilege levels, which aims at eliminating malicious contention and reuse.
One approach is providing additional hardware resources to back up (certain portions of) predictor table upon context switches~\cite{dhodapkar2001saving}.
And backed up table contents will be recovered for the thread swapped in.
Nevertheless, there exist two limits:

\ding{172} Physical isolation is likely to incur non-trivial resource overhead.
A milder version of physical isolation could be part of a solution with acceptable overheads.
For instance, BRB strives to keep a small (1-3KB) portion of the state backed up~\cite{vougioukas2019brb}.
\ding{173} Constrained by hardware overhead, the number of back up tables is limited.
It is impossible for every thread to have private history table.
Then there may be attacks in subsequent software programs using the same back up table.

\subsection{Our design philosophy}
\label{philosophy}
Both flush-based logical and physical isolation mechanisms have their own
design space. A solution needs to satisfy the following criteria to achieve a
secure and yet practical branch predictor design:

\begin{itemize}
    \item \emph{Enabling isolation between different processes and different privilege spaces;}
    \item \emph{Versatile to accommodate multiple branch predictors;}
    \item  \emph{Lightweight implementation to minimize changes to existing branch predictors;}
    \item  \emph{Tolerance of interference between SMT hardware threads.}
\end{itemize}

Instead of isolating based on the flush or separated resources, we want to achieve isolation of the \emph{content} in these
hardware tables with some lightweight processing using randomization as
follows. 
\begin{itemize}
\item{\textbf{Content encoding:}}. We propose to use hardware-based
thread-private random numbers to encode the contents of the branch predictor
tables (both direction and destination histories). Specifically, the data
is encoded (think xor for now) with the key before written into the table and
decoded after read from the table. This process is similar to the use of a
\emph{key} to encrypt data, where the key is a hardware-generated
thread-private random number that changes periodically (upon context switch or privilege changes). Such a mechanism
obfuscates the information adding difficulties to cross-process or
cross-privilege analysis and perception. It achieves a similar protection effect of
\emph{Precise Flush}, but adds little in terms of space or time overheads.

\item{\textbf{Index encoding:}} In addition to entry encoding, we also use
randomization of indexes, which sets up additional obstacles for an attacker in 
all three steps of locating, priming, and probing the predictor.
Index encoding also requires little change
to conventional branch predictor designs and incurs low overheads.
\end{itemize}

% that's all folks

\begin{figure*}[tb]
\centering
  \includegraphics[width=0.98\textwidth]{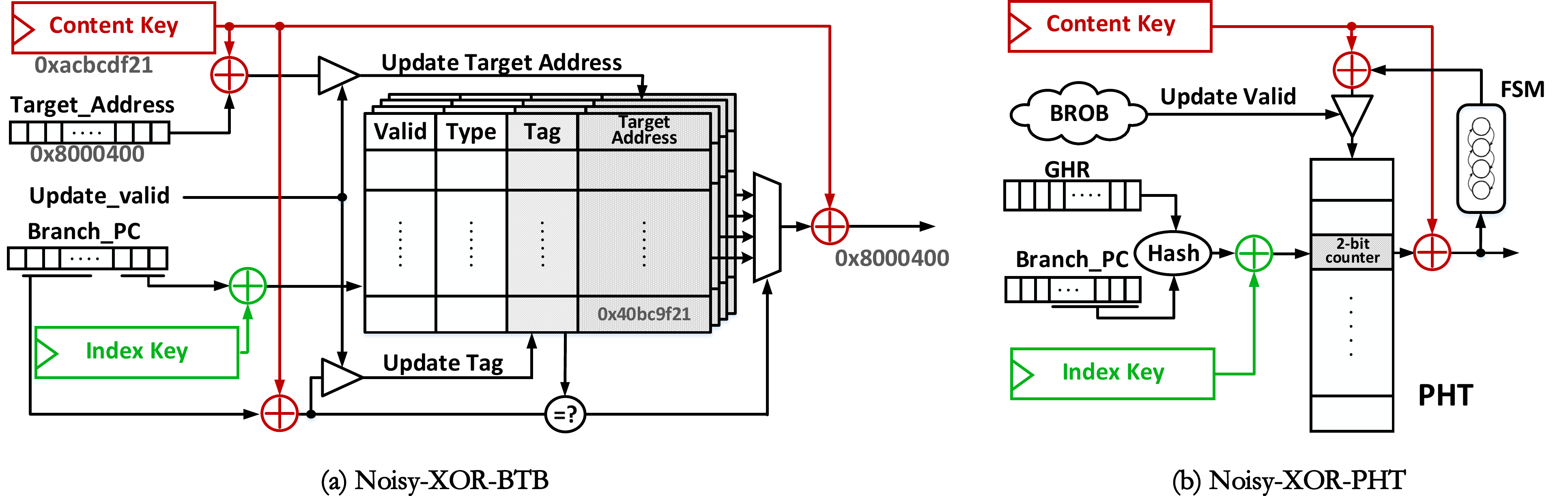}
\caption{Microarchitecture implementation of content encoding and index encoding. 
The red modules are designed for XOR-BP; the combination of red the green ones are for Noisy-XOR-BP. 
We take Gshare architecture as the example to describe our design for PHT.}
\label{Noisy_XOR_BP}
\vspace{-1.0em}
\end{figure*}

\section{Lightweight Content Isolation Mechanisms}
\label{XOR}

The general idea is straightforward: we transform both index and table content
with thread-private keys; these keys change under certain conditions.
We now discuss the implementation issues. We start with the example of BTB using 
the xor operation as the encoding/decoding (\emph{XOR-BTB}).

\subsection{Implementation of XOR-BTB}

Each active hardware thread context will be allocated a
thread-private random number as the \emph{key} to encode or decode information
stored in the table, such as tag and target address. The simplest coding operation is XOR. 
(We will see later that this can be exchanged for stronger isolation.)

\textbf{Lookup:} When predicting the target of a branch,
the processor usually employs partial PC bits 
to index the BTB and compares the most significant bits of the PC
with the tag of each way. If a match is found, the target address saved in
the entry will be taken out as the target of the branch. In
\emph{XOR-BTB},  the higher bits of the PC are XORed with
the \emph{Content Key} for tag match (Figure~\ref{Noisy_XOR_BP}~a). Finally, the stored target address (0x40bc9f21)
is XORed with the Content Key to produce the actual predicted target (0x80004000).
These additional steps are shown in red in the figure.

When a different thread ($i$ with key $k_i$) executes a branch, 
it will not obtain the original tag or target address updated by thread $j$
due to the difference in their keys. This provides a logical 
isolation of the content of BTB among different threads.

\textbf{Update:} In a typical out-of-order processor, when an indirect branch 
instruction reaches the stage of execution, the actual destination will be 
compared to the predicted address. If they are different, it is considered as a BTB
misprediction and the corresponding content in the BTB needs to be updated.
In \emph{XOR-BTB}, both the tag and the target address are encoded (XORed with the
thread's content key) before saved in the BTB. In Figure~\ref{Noisy_XOR_BP}~(a), 
the actual target address 0x80004000 is XORed  with the current content key
(0xacbcdf21) to produce the \emph{encrypted} address 0x40bc9f21, which is
stored in the BTB. 

Note that in \emph{XOR-BTB}, the tag of BTB is also encoded lest an attacker
could use performance counters as a covert channel to sense possible resource
contention~\cite{evtyushkin2015covert}. For instance, Intel processors
provide BTB-related performance counters~\cite{Intel2017perf}, enabling an
attacker to observe the case which hits the BTB but has a misprediction due to
incorrect address.

\subsection{Extending to other tables or predictors}
\label{enhanced}

\begin{figure}[!t]
\centering
  \includegraphics[width=0.48\textwidth]{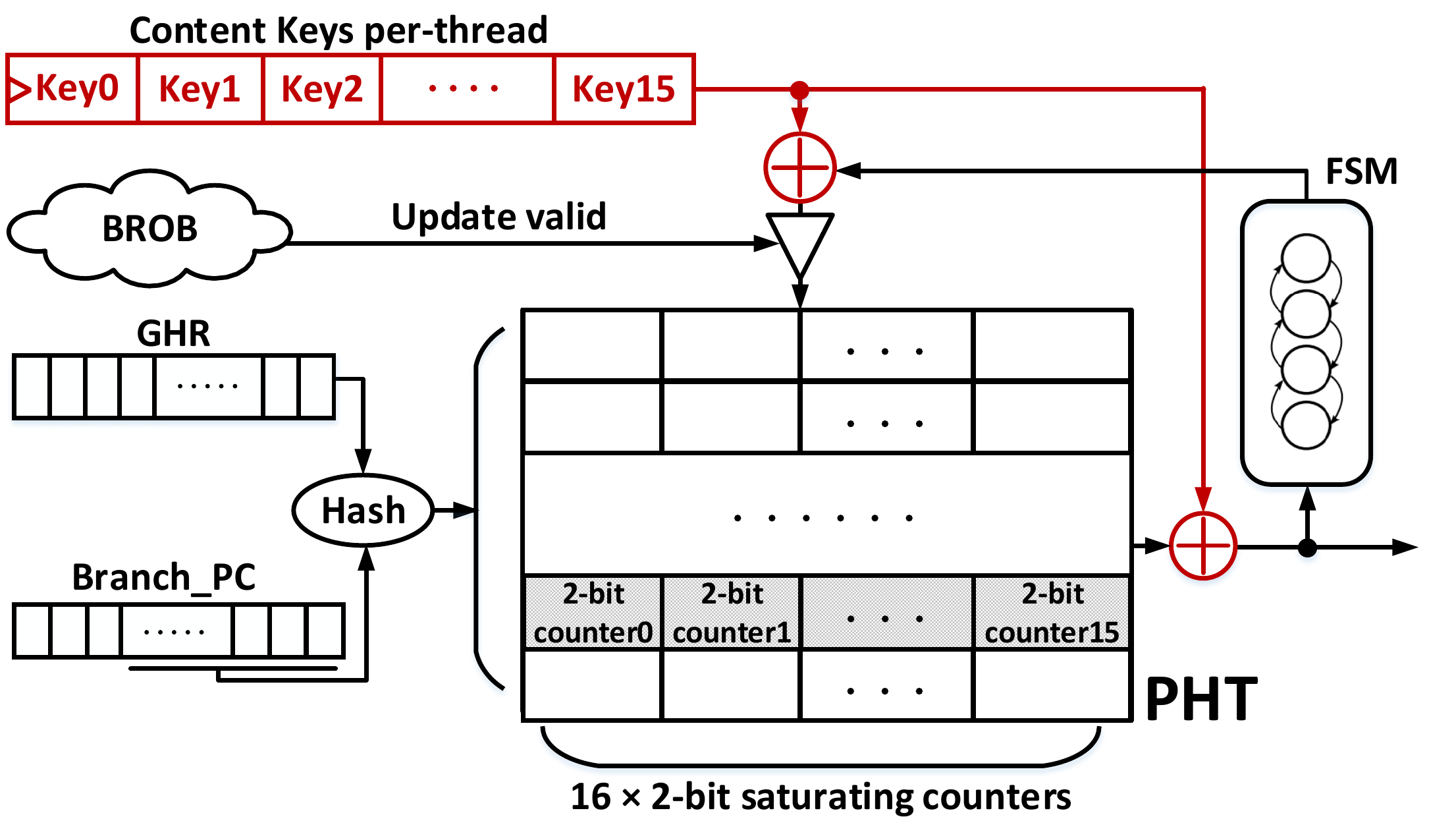}
\caption{Microarchitecture implementation of \emph{Enhanced-XOR-PHT}.}
\label{enhanced_XOR_PHT_arch}
\vspace{-1.3em}
\end{figure}

The general idea discussed above is applicable to any table structure in
a branch predictor regardless of the specific algorithm involved and the
detailed organization of the underlying tables. Also note that the encoding
and decoding operations need not even be done on a single logical entry of
the table. For instance, in a simple 2-bit PHT, a
logical entry contains only two bits. Using XOR on these two bits (\emph{XOR-PHT}) may not
yield sufficient obfuscation (Figure~\ref{Noisy_XOR_BP}(b)).
Instead, the encoding and decoding can operate on word basis of an arbitrary 
length, regardless of the logical meaning of bits (\emph{Enhanced-XOR-PHT}).
For instance, shown in Figure ~\ref{enhanced_XOR_PHT_arch}, ~a 4K-entry PHT with 2-bits per entry can be considered as a 
256-entry array of 32-bit words. We could then encode and decode these 32-bits word
using 32-bit keys. Indeed, the physical implementation of the table using
SRAM is most likely using a wider row already. An alternative view of this
issue is that different logical entries nearby in the PHT can use different keys for
encoding/decoding.

\textbf{Update:}
After the branch instruction is committed, the saturating counter needs to be updated, whether the results of prediction is correct or not.
There are two typical ways for such an update depending whether a \emph{branch reorder 
buffer} (BROB) is used to store the counter value.

\begin{itemize}

\item If a BROB is used, we take the counter value from the buffer, update it,
encode the updated value with the key, and store it to PHT;

\item If not, the original counter needs to be read out of the PHT (and decoded) first
before being updated, re-encoded, and written back.

\end{itemize}

\subsection{Randomized index}
\label{Noisy-XOR-BP}

The content randomization mechanism encrypts branch history contents,
but the history information still resides at predictable locations in the
tables. We propose to add index randomization (\emph{Noisy-XOR-BP}) which can further
increase the noise and help thwart attacks like contention-based ones.
As discussed earlier in
Section~\ref{vulnerabilities}, an attacker would first locate the entry
corresponding to the branch predictor (Locate phase) of the target branch
instruction, and then train or analyze the corresponding entry at the Prime
and Probe phases. An important prerequisite for this process is that an
attacker can locate the entry based on the address of the branch instruction.
That is, the indexing mechanism of the existing branch predictor is relatively
fixed, which gives an attacker the opportunity to pinpoint.

The purpose of the \emph{Noisy-XOR-BP} mechanism is to break the simple and fixed
indexing mechanism of existing branch predictors. Like \emph{XOR-BP}, \emph{Noisy-XOR-BP}
dynamically assigns another random index key private to each thread. The index
key is XORed with the lower part of the PC to generate the index whenever there is
any table lookup (green part in Figure~\ref{Noisy_XOR_BP}). 
Since only the encoded index is used for actual table lookup, a
contention-based attack will only obtain information of the \emph{encoded} index.

Note that this index key must be updated in a timely manner. Otherwise, an attacker
would still have the opportunity to eventually construct branches which
shares the same index as the target branch of victim.
In the subsequent description, both \emph{Noisy-XOR-BTB} and \emph{Noisy-XOR-PHT} 
include content and index encoding.
In particular, \emph{Noisy-XOR-PHT} encodes content with \emph{Enhanced-XOR-PHT} mechanism.

In terms of the thread-private random number, the generation, updating, and
exception handling are the same with that used in content randomization. In
practice, the hardware random number generator can generate a single random number
whose different (possibly overlapping) portions are used as keys in content
and index randomization.

\subsection{Implementation issues}

The isolation of content in our design depends on the security of the key. We
need to prevent information leakage not only between different programs, but
also between different privilege levels (such as user mode and kernel mode)
when running the same program.
Thus when a new thread is switched in or when the thread's privilege level
changes, we change the key as follows: We use a dedicated hardware register
per hardware thread to record the key. Such a thread private register is
invisible to software. Once a context switch or a privilege switch occurs, a
new random number will be generated and updated to this private register.
All subsequent accesses would use the updated key. Correspondingly, OS and
hypervisor also have their own keys. Since there are many mature methods for
hardware random generation~\cite{intel2014drng,liberty2013truerng}, we assume
these random numbers can be generated using a dedicated hardware mechanism.

Now consider the SBPA attack on BTB as an example. The attacker first primes
target addresses of a certain set in BTB. These address are XORed with the
current private key of the attacker before stored in BTB. The attacker then
waits for context switch to allow the victim execution to leave footprint in
the prime set before context-switched back for the Probe stage. At this point,
the private key has be automatically changed by the hardware. The result is
that the attack will sense misses in the BTB for all the primed addresses
regardless of whether the victim evicted any entry in the set.

Finally, note that while we use XOR throughout the discussion, the only
requirement for the encoding operation is that they are easily reversible
so that both encode and decode operations are lightweight enough to not cause
critical path timing problems. Adding shifting and/or scrambling in the process,
or using small lookup tables are all possible options.

\subsection{Security analysis}

\begin{scriptsize}
\begin{table}[!t]
%\begin{table}[!t]
  \caption{Security comparison.}
  \label{security_table}
\setlength{\abovecaptionskip}{0.cm}
  \centering
  \small
  \scriptsize
  \begingroup
  \setlength{\tabcolsep}{4.3pt}
  \linespread{1.3}
  \begin{threeparttable}
  \begin{tabular}{|l|c|c|c|c|c|}
  \hline
  \multicolumn{2}{|c|}{\multirow{2}{*}{\textbf{Defense Mechanism}}} & \multicolumn{2}{c|}{\textbf{Single-threaded core}} & \multicolumn{2}{c|}{\textbf{SMT core}}\\
  \cline{3-6}
  \multicolumn{2}{|c|}{} & \textbf{Reuse} & \textbf{Contention} & \textbf{Reuse} & \textbf{Contention} \\
  \hline
  \multirow{4}{*}{\rotatebox{90}{\textbf{BTB}}} & Complete Flush & \emph{Defend} & \emph{Defend} & \emph{No Protection} & \emph{No Protection} \\
  \cline{2-6}
  & Precise Flush & \emph{Defend} & \emph{Defend} & \emph{Defend}\tnote{1} & \emph{No Protection} \\
  \cline{2-6}
  & XOR-BTB & \emph{Defend} & \emph{Defend} & \emph{Mitigate} & \emph{No Protection} \\
  \cline{2-6}
  & Noisy-XOR-BTB & \emph{Defend} & \emph{Defend} & \emph{Defend} & \emph{Mitigate} \\
  \hline
  \multirow{5}{*}{\rotatebox{90}{\textbf{PHT}}} & Complete Flush & \emph{Defend} & \emph{Defend} & \emph{No Protection} & \emph{Defend} \\
  \cline{2-6}
  & Precise Flush & \emph{Defend} & \emph{Defend} & \emph{Defend}  & \emph{No Protection}\tnote{2} \\
  \cline{2-6}
  & XOR-PHT & \emph{Mitigate} & \emph{Defend} & \emph{No Protection} & \emph{Defend} \\
  \cline{2-6}
  & Enhanced-XOR-PHT & \emph{Defend} & \emph{Defend} & \emph{Mitigate} & \emph{Defend} \\
  \cline{2-6}
  & Noisy-XOR-PHT & \emph{Defend} & \emph{Defend} & \emph{Mitigate} & \emph{Defend} \\
  \hline
  \end{tabular}
  \begin{tablenotes}
  \scriptsize
  \item[1] \emph{Precise Flush} needs thread ID, with which branches in different hardware thread cannot use the others' history.
  \item[2] Adding thread ID brings significant hardware overhead for PHT. Besides, there are contention based attacks on SMT.
  \end{tablenotes}
%   \vspace{-1em}
  \end{threeparttable}
  \endgroup

\vspace{-2em}
\end{table}
\end{scriptsize}

Table ~\ref{security_table} summarizes the security of different isolation mechanisms.
\emph{Complete Flush} and \emph{Precise Flush} represent two costly and rather
impractical mechanisms. Yet, because flushing only happens during context
or privilege switches, they still fail to protect against certain attacks
in SMT cores.
In contrast, XOR-based isolation mechanisms are both more secure and more
light-weight than these flush-based mechanisms. The specific analysis is
detailed as below.

\begin{enumerate}
\item[(1)] Security of \emph{Noisy-XOR-BTB}
\end{enumerate}

In this analysis, we assume a BTB is \emph{W}-way set associative  with \emph{S}-bit set index and \emph{T}-bit tag per entry.
Three major scenarios are analyzed.

\textbf{Scenario 1: Reuse based attacks on single-threaded and SMT cores can be defended.}
For a reuse-based attack to work, the entry left by one party is being
translated in multiple ways before being used by another party, which makes it
exceedingly difficult to maintain controlled manipulation. Take malicious
training for example. The attacker wants to lay traps in the BTB to direct the
victim to a meaningful location.
First, these entries need to result in a BTB hit to be even considered. For
that the (partial) tag left by the attacker has to match the encoded tag of
a victim branch. The chance for one entry to have a BTB hit is $1/2^T$. The
attacker can certainly lay many such traps, increasing the overall probability
to some degree. But there is a second hurdle to clear: The
content of the entry will need to lead to a meaningful location, one that contains
the malicious code.  Since the content is XORed with another unknown key,
the probability of leading the victim to a specific address is $1/2^N$, N being
the number of address bits. Again, the attacker can certainly prepare many
such traps and extend the attack over many intervals. But overall, the
chance of success is against a very large denominator $(2^{N+T})$.

The general analysis applies to the case of an SMT core as well. The slight 
advantage to an attacker there is that they can continue to re-plant new 
traps in the BTB, whereas in a single-threaded core, those entries are gradually
evicted and the attack strength reduces with time in a context switch interval.

\textbf{Scenario 2: Contention based attacks on single-threaded cores can be defended.}
Since there are context switches between prime phase and probe phase, a thread's 
own history in the previous phase becomes unrecognizable in the latter phase.
Thus the attacker cannot sense the conflict caused by the target branch.

\textbf{Scenario 3: Contention based attacks on SMT cores can be mitigated.}
In a contention-based attack carried out on a conventional SMT core, an attack like Jump~\cite{evtyushkin2016jump} can observe evictions of its own entry in BTB and thus infer the address of taken branch executed by the victim. In our system,  different hardware thread has different private key for indexing. Without the key of the victim thread, the attacker can no longer make the inference of the branch address.  

However, it is conceivable to extend certain attacks such as SBPA as follows. The attacker can prime the entire BTB with its own entries and  invoke single-step mode to force the victim to execute a single branch of interest. Consequently, the mere change in any BTB content indicates a BTB update due to a taken branch.  In such a hypothetical case, no encryption of content or the index of BTB can be of any help as  the attacker is only sensing the fact there is an update to BTB, without any interest in the content or index of the update. We note that this type of attack should be handled differently altogether and is rather beyond the scope of this paper. We only highlight one possible approach here to suggest that it also can be thwarted. The attack described above reduces the demand on the information content from the BTB (or PHT) but relies on highly precise control of victim's execution (a single branch). A reasonable counter measure is for the system to detect extreme reduction of execution speed, and subsequently bypass update of any microarchitectural resources completely as these updates are unlikely to matter for execution speed.  

\begin{figure}[!t]
\setlength{\abovecaptionskip}{-2em}
\begin{lstlisting}[language={[ANSI]C},
        numbers=left,
        xleftmargin=0.5em,
        xrightmargin=0.5em,
        basicstyle=\scriptsize,
        tabsize=2,
        % backgroundcolor=\color{red!15},
        frame=single,
        % framerule=1pt,
        rulesep=1pt,
        % rulesepcolor=\color{red},
        % rulecolor=\color{red},
        ]
/* Shared pointer and function */
static void (*p)();  // A function pointer shared by
        // the attacker thread and the victim thread
void shared_interface(){
  p();  // Execution history is recorded in BTB
}
 -----------------------------------------------------
/* Attacker Thread */
{ // p points to attacker_function() in attacker thread
  shared_interface();  // Train to execute attacker's p
  clflush(p);
  clflush(&side_line); // FLUSH side_line
  sleep(1);      // Switch to victim thread in Line 22
  a=side_line;   // Assess the time to RELOAD side_line
}
void attacker_function(){
    a=side_line; // Leave observable traces
}
 -----------------------------------------------------
/* Victim Thread */
{   // p points to victim_function() in victim thread
  shared_interface();
  ...
}
void victim_function(){
  a=sec;
}
\end{lstlisting}
\captionof{codetype}{PoC Code piece of BTB attacks.}
\label{btb_poc}
\end{figure}
% }
    
\begin{enumerate}
\item[(2)] Security of \emph{Noisy-XOR-PHT}
\end{enumerate}

As mentioned in Section ~\ref{vulnerabilities}, there is no contention based attack 
in PHT because a branch history updates the older history, rather than eviction.
Security for two typical scenarios is analyzed as following.

\textbf{Scenario 4: Reuse based attacks on single-threaded cores can be defended.}
First, the attacker cannot manipulate the victim deterministically to execute the wrong path speculatively. 
Because the private key changes at each context switch, the attacker cannot
train the status to any deterministic direction for the target victim thread.
Second, \emph{Noisy-XOR-PHT} poses strict requirements on attacks to perceive the direction of target branch.
An attacker can no longer infer the direction through one Prime-Probe operation.

There is one corner case: if an attacker can find a reference branch which
employs the same private key as the target branch, and if this reference
branch's direction is easily known (e.g., it is biased), then the attacker
can indirectly infer the target branch's direction. Finding such a reference
branch is easy if we use a simple XOR scheme with a fixed key width. The
root cause is the fixed mapping relationship between the branch instruction
address and content keys index within a thread. Breaking this fixed mapping
relationship can be done in a number of relatively simple ways while still
allowing easy decoding for normal use. For instance, the index can be selected
from PC dynamically and shifted based on the index key.

\textbf{Scenario 5: Reuse based attacks on SMT cores can be mitigated.}
In \emph{XOR-PHT}, an attacker may obtain sensitive information
from analyzing the update sequences from target branch only. With
\emph{Noisy-XOR-PHT}, the attacker needs to traverse every entry.  This traversal
prolongs Prime-Probe operation, resulting in the decrease
of information leakage bandwidth. Furthermore, signals,
interrupts, and exceptions are used for high resolution in existing
attacks~\cite{evtyushkin2016jump,lee2017inferring,evtyushkin2018branchscope}.
All of them involve kernel operation and thus the change of private key, which defends attacks effectively.

\begin{figure}[!t]
\setlength{\abovecaptionskip}{-2em}
\begin{lstlisting}[language={[ANSI]C},
        numbers=left,
        xleftmargin=0.5em,
        xrightmargin=0.5em,
        basicstyle=\scriptsize,
        tabsize=2,
        % backgroundcolor=\color{red!15},
        frame=single,
        % framerule=1pt,
        rulesep=1pt,
        % rulesepcolor=\color{red},
        % rulecolor=\color{red},
        ]
/* Shared function */
void shared_interface(int i){
  if(i<array_size) // Accessible to attacker and victim
    a=sec;
  else
    a=side_line;   // Leave observable traces
}
 -----------------------------------------------------
/* Attacker Thread */
{
  for(int i=array_size;i<2000+array_size;i++){
    shared_interface(i); // Train Line 3 to Not Taken
  }
  
  clflush(&array_size);
  clflush(&side_line);   // FLUSH side_line
  
  sleep(1);     // Switch to victim thread to Line 24
  a=side_line; // Assess the time to RELOAD side_line
}
 -----------------------------------------------------
/* Victim Thread */
{
  ...
  shared_interface(x);
  ...
}
\end{lstlisting}
\captionof{codetype}{PoC Code piece of PHT attacks.}
\label{pht_poc}
\end{figure}

\begin{figure*}[htb]
\centering
  \includegraphics[width=0.98\textwidth]{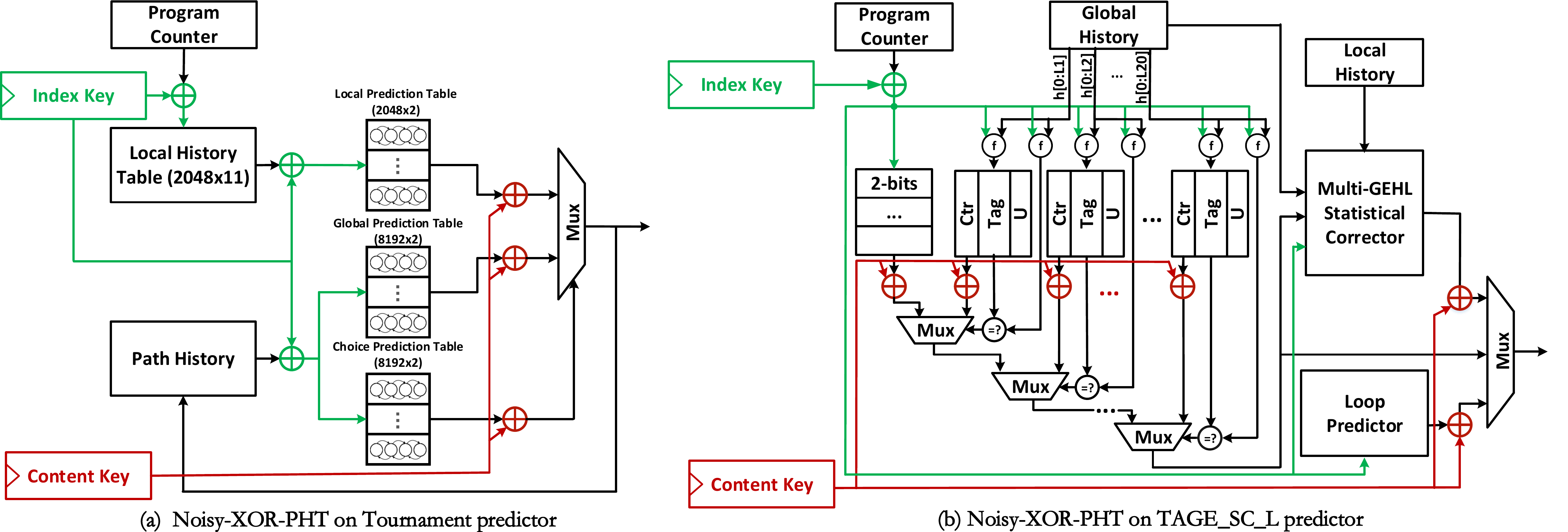}
\caption{
Implementation of \emph{Noisy-XOR-PHT} on two example predictors. All the
tables here use the shared index key and content key, and of course each table
can also have their own index key and content key. {\bf (a) Tournament:} 
the first level holds 11 bits of branch pattern history for up
to 2048 branches. This 10-bit pattern picks from one of 2048 prediction
counters. The global predictor is an 8192-entry table of 2-bit saturating
counters indexed by the path (or global) history of the last 12 branches.
The choice prediction, or chooser, is also an 8192-entry table of 2-bit
prediction counters indexed by the path history. 
{\bf (b) TAGE\_SC\_L:} the TAGE component consists of a base predictor (16Kbits prediction, 4Kbits
hysteresis), two bank-interleaved tagged tables featuring respectively ten 12-bit, 1K-entry banks,
and twenty 16-bit, 1K-entry banks. The respective tag widths are 8 and 11
bits, a 3000-bit global history length, a 27-bit global path length, sixteen
5-bit USEALT counters, a 19-bit counter for monitoring the allocation policy.
The loop predictor features 256 entries and is 4-way associative (256 x 52 bits).
The Multi-GEHL statistical corrector consists of seven GEHL-like components respectively indexed using the global conditional branch history (two tables), the global history of the backward branches (two tables), an IMLI counter indexed table, another IMLI-based GEHL predictor and three local history GEHL components respectively with 256-entry, 16-entry, and 16-entry local history.
}
\label{implementation_Tournament}
\vspace{-1.0em}
\end{figure*}

\begin{enumerate}
\item[(3)] Attack \& defense experiments
\end{enumerate}
To evaluate the effectiveness of our mechanism, we conduct experiments
with proof-of-concept (PoC) attacks for BTB and PHT respectively on
our FPGA-based processor prototype (configuration is introduced in Section
~\ref{evaluation}). The PoC codes (detailed in Listing~\ref{btb_poc} and \ref{pht_poc})
proceed in a realistic scenario: cross-thread, within the same address domain.

In our experiment, we repeat the attack 10000 iterations. In the PHT attack,
we consider 100 attempts of training as one iteration. A successful attack
means that the victim branch jumps to the trained direction more than 90
times. For the baseline processor without any defense mechanism, the accuracy
of training BTB and PHT is 96.5\% and 97.2\%, respectively. The accuracy on
an Intel E5-2697 v4 processor is higher than 99.9\%. With \textbf{XOR-based
Isolation}, the accuracy of training both BTB and PHT decreases to less than
1\%.\footnote{The 1\% apparently successful attacks stem from the limitations of the
RISCV experimental platform and software noises. For example, we determine the
success of attack by observing Flush+Reload cache side channels. However,
flushing a cache line precisely is not supported in RISCV instruction set,
so we employ evicting the whole cache with large arrays. This presents
false positive measurement noises on successful attack. 
However, an adversary cannot exploit these noises to construct
attacks.}
In summary, our mechanism introduces effective
protection against these attacks.
\section{Evaluation}
\label{evaluation}

\subsection{Methodology}

\begin{scriptsize}
\begin{table}[!t]
  \caption{OoO Processor Core Configurations.}
  \label{boom_configuration}
\setlength{\abovecaptionskip}{0.cm}
\setlength{\belowcaptionskip}{-0.25cm}
  \centering
  \small
   \begin{tabular}{|p{1.8cm}|p{2.8cm}|p{2.8cm}|}
%   \begin{tabular}{|c|c|}
    \hline
    \multirow{2}{*}{\textbf{Parameter}} & \multicolumn{2}{c|}{\textbf{Configurations}} \\
    \cline{2-3}
    & \multicolumn{1}{c|}{\textbf{FPGA prototype}} & \multicolumn{1}{c|}{\textbf{Gem5 simulation}} \\
    
    \hline
    \hline
    ISA & RISC-V & ALPHA\\
    \hline
    \multirow{2}{*}{Frequency} & 2\emph{GHz} & \multirow{2}{*}{2.5\emph{GHz}} \\
    & (FPGA runs at 50\emph{MHz}) & \\
    \hline
    \multirow{2}{*}{Processor type} & 4-decode,4-issue, &  8-decode,8-issue, \\
    & 4-commit & 8-commit \\
    \hline
    Pipeline depth & 10 stages & 19 stages\\
    \hline
    ROB/LDQ/STQ & 64/16/16 entries & 352/128/72 entries\\
    \hline
    Issue Queue & 20/16/10 (mem/int/flt) & 120 \\
    \hline
    BTB & 256 $\times$ 2-way & 1024 $\times$ 4-way \\
    \hline
    \multirow{4}{*}{PHT} & TAGE: 33 KB & TAGE\_SC\_L: 66.6KB \\
    & 6 $\times$ 4096 entries & or LTAGE: 32KB \\
    & history length: & or Tournament: 6.3KB \\
    & 12, 27, 44, 63, 90, 130 & or Gshare: 2KB\\
    \hline
    ITLB/DTLB & 8/8 entries & 64/64 entries\\
    \hline
    L1 ICache & 32KB, 8-way, 64B line & 32KB, 4-way, 64B line\\
    \hline
    L1 DCache & 32KB, 8-way, 64B line & 48KB, 4-way, 64B line\\
    \hline
    L2 Cache & 1MB, 16-way, 64B line & 512KB, 16-way,64B line\\
    \hline
    L3 Cache & None & 4MB, 32-way, 64B line\\
    \hline
  \end{tabular}
\vspace{-1.5em}
\end{table}
\end{scriptsize}

We built an FPGA prototype based on the open-source Berkeley Out-of-Order
RISC-V processor (BOOM) as the main platform for experimental
analyses~\cite{AsanovicBOOM,WatermanRISCV}. The main parameters of
this system are shown in Table~\ref{boom_configuration}. On this
platform, we implemented the \emph{XOR-BP} and \emph{Noisy-XOR-BP}
mechanisms to evaluate the performance impacts. Since our processor
prototype does not yet support SMT or some branch predictors,
we also modeled an out-of-order SMT processor using the cycle-level Gem5
simulator~\cite{binkert2011gem5}. 
This SMT core is modeled after the latest Intel Sunny Cove core \cite{SunnyCove} as shown in Table ~\ref{boom_configuration}. 
In addition to Gshare, we experimented with three different
branch predictors: Tournament~\cite{kessler1999the}, LTAGE~\cite{seznec_LTAGE}, and TAGE\_SC\_L~\cite{seznec_TAGE_SC_L}. The modifications of Tournament and TAGE\_SC\_L predictor with \emph{Noisy-XOR-PHT} are shown
in Figure~\ref{implementation_Tournament}. 
In configuring TAGE\_SC\_L, we increased the size of the base and the loop predictor moderately as that improves prediction accuracy in our benchmarks.

On the RISC-V based FPGA platform, we randomly selected 12 combinations
(\emph{case1} to \emph{case12} for the single-threaded core in Table~\ref{benchmarks})
from SPEC CPU 2006 benchmark suite~\cite{henning2006spec}, each consisting of a target benchmark (first in the entry)
and a background benchmark (second in the entry) to build the scenario of context switch. Using the
\emph{train} input set, we run these combinations on the Linux operating
system to evaluate the performance by measuring the execution time of target
benchmark. Our Linux operating system uses the default context switching
frequency of 250Hz.

For the simulated SMT-2 platform, we randomly selected 12 pairs 
of applications  from the SPEC CPU 2006 benchmark suite 
(column SMT-2 in Table~\ref{benchmarks}) to run concurrently.
These benchmarks run
in the mode of System Call Emulation (SE) with the typical context-switching frequency in Linux. Two billion instructions are used
for warm-up, and then we count the execution cycles of the next two billion
instructions executed by either thread.

Experimental setups with different isolation mechanisms are named as
follows:

\textbf{\emph{Baseline}}: The original OoO processor using branch predictor without
any isolation mechanism.

\textbf{\emph{IsolationMechanism-nM}}: The name of corresponding isolation mechanism
is followed by the period of context
switch in cycles due to timer interrupt. The standard Linux switches context every 
4 milliseconds (or  8 million 2GHz CPU cycles). For
example, XOR-BP-8M represents \emph{XOR-BP} when the
thread is switched every 8 million CPU cycles.

\textbf{\emph{PredictorName-FlushMechanism}}: \emph{PredictorName} represents
the corresponding combination of branch predictor and flush
mechanism, which includes CF for \emph{Complete Flush} and PF for \emph{Precise Flush}. For example, Gshare-CF
represents a processor using Gshare branch predictor and \emph{Complete Flush}
isolation mechanism.

\begin{scriptsize}
\begin{table}[!pt]
  \caption{Benchmark sets.}
  \label{benchmarks}
\setlength{\abovecaptionskip}{0.cm}
\setlength{\belowcaptionskip}{-0.25cm}
  \centering
  \small
%   \begin{tabular}{|c|c|c|c|}
  \begin{tabular}{|p{2.3cm}|p{3cm}|p{3cm}|}
    \hline
    \multicolumn{1}{|c}{\textbf{Test Number}} & \multicolumn{1}{|c}{\textbf{Single-threaded core}} & \multicolumn{1}{|c|}{\textbf{SMT-2}}\\
    \hline
    \hline
    \multicolumn{1}{|c|}{case1} & gcc+calculix & zeusmp+lbm \\ 
    \hline
    \multicolumn{1}{|c|}{case2} & milc+povray & zeusmp+dealII \\
    \hline
    \multicolumn{1}{|c|}{case3} & bzip2\_source+soplex & bwaves+milc \\
    \hline
    \multicolumn{1}{|c|}{case4} & namd+sphinx3 & leslie3d+gromacs \\
        \hline
    \multicolumn{1}{|c|}{case5} & hmmer+GemsFDTD & dealII+sjeng \\
        \hline
    \multicolumn{1}{|c|}{case6} & gobmk+libquantum & gromacs+astar \\
        \hline
    \multicolumn{1}{|c|}{case7} & gromacs+GemsFDTD  & gobmk+h264ref \\
        \hline
    \multicolumn{1}{|c|}{case8} & mcf+astar & libquantum+milc \\
        \hline
    \multicolumn{1}{|c|}{case9} & soplex+hmmer & gobmk+gromacs \\
        \hline
    \multicolumn{1}{|c|}{case10} & libquantum+calculix & milc+bzip2\_source \\
        \hline
    \multicolumn{1}{|c|}{case11} & mcf+perlbench & libquantum+omnetpp \\
        \hline
    \multicolumn{1}{|c|}{case12} & bwaves+namd & zeusmp+gobmk \\
        \hline
  \end{tabular}
\vspace{-1.3em}
\end{table}
\end{scriptsize}

\subsection{Evaluation on FPGA-based single-threaded RISC-V core}

\subsubsection{XOR-BP performance impacts}

When an application resumes its execution
after being swapped out and back again on \emph{Baseline}, it will benefit from its residual
state in the BTB. With \emph{XOR-BTB} mechanism, however, each context switch results in
a change of the content key. The branch predictor is thus unable to correctly
decode any residual state. This in general leads to a performance loss
relative to \emph{Baseline}. Keep in mind, though, that in some special cases,
the loss of residual state can actually improve performance as we will
see later. Below, we evaluate the performance impacts of \emph{XOR-BTB}, 
\emph{XOR-PHT}, and their combination at different context switching frequencies.
 
\begin{figure}[!t]
\centering
  \includegraphics[width=0.48\textwidth]{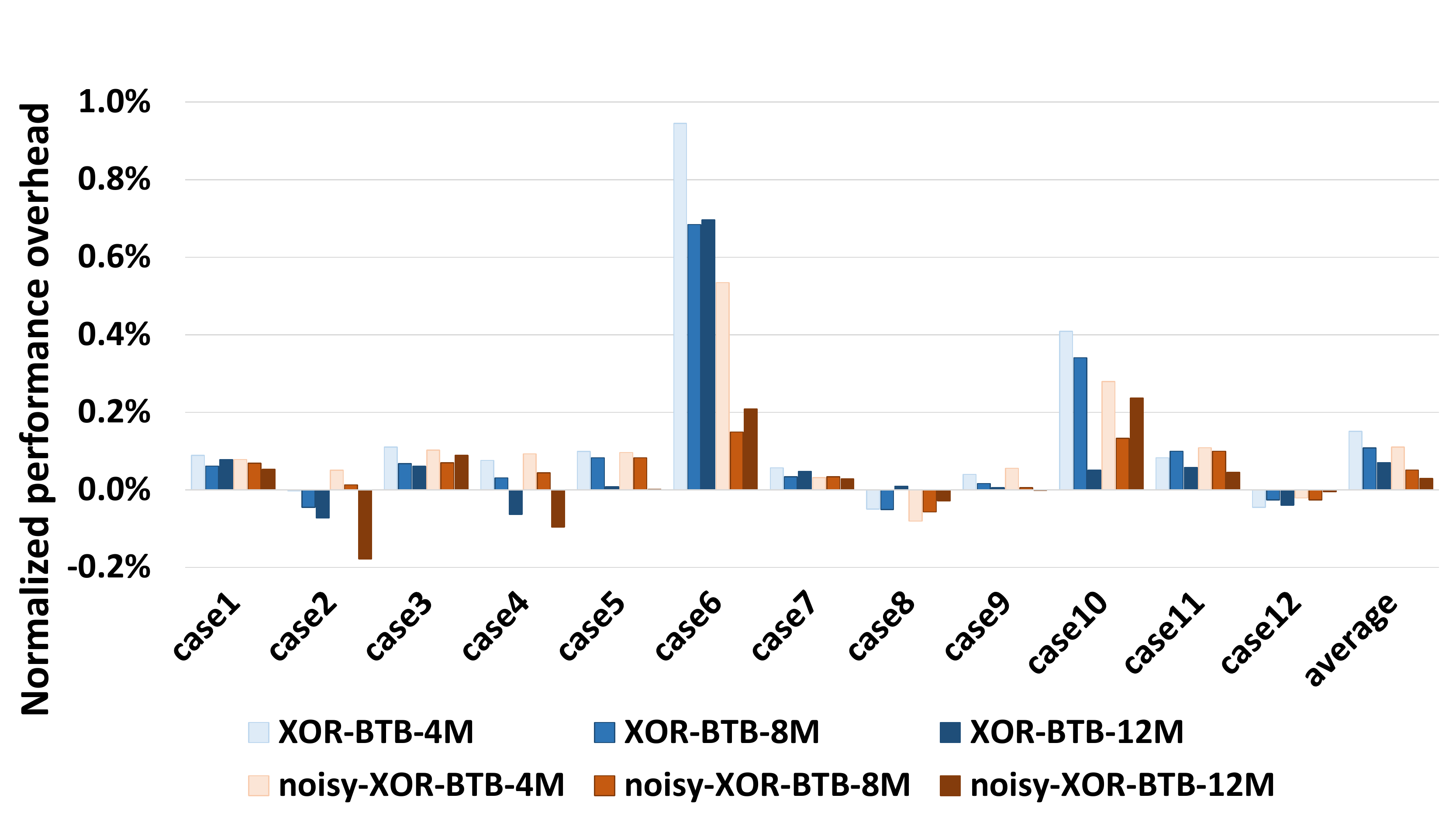}
\caption{Performance overhead of \emph{XOR-BTB} and \emph{Noisy-XOR-BTB}.}
\label{XOR_BTB}
\vspace{-1.4em}
\end{figure}

As shown in Figure~\ref{XOR_BTB}, compared with \emph{Baseline}, the average
performance loss of \emph{XOR-BTB} is less than 0.2\%. Noisy-XOR-BTB, which
adds randomized index, introduces no additional performance loss. The highest
performance loss can approach 1.0\% from case6 (combination of gobmk and
libquantum). For the pair of gobmk and libquantum, there are many more
residual BTB entries (between 500 and  800 entries)
upon switching back -- compared to, say, the namd and sphinx3 pair of 30-300 entries.
And these entries contributed to the high prediction accuracy of BTB, which 
reaches 85.2\% and 99.3\% while running on \emph{Baseline}.
For such cases, the effect of flushing BTB (with key change) usually results in
a noticeable performance penalty.

For case2 (the combination of milc and povray), flushing the BTB has the
unusual effect of actually improving the execution speed. It turns out,
while the miss rate of BTB increases due to flushing, it helps to overturn some
incorrect \emph{direction} predictions as the processor in our implementation
simply reverts to fall-through prediction when the target is unavailable.

\begin{figure}[!t]
\centering
  \includegraphics[width=0.48\textwidth]{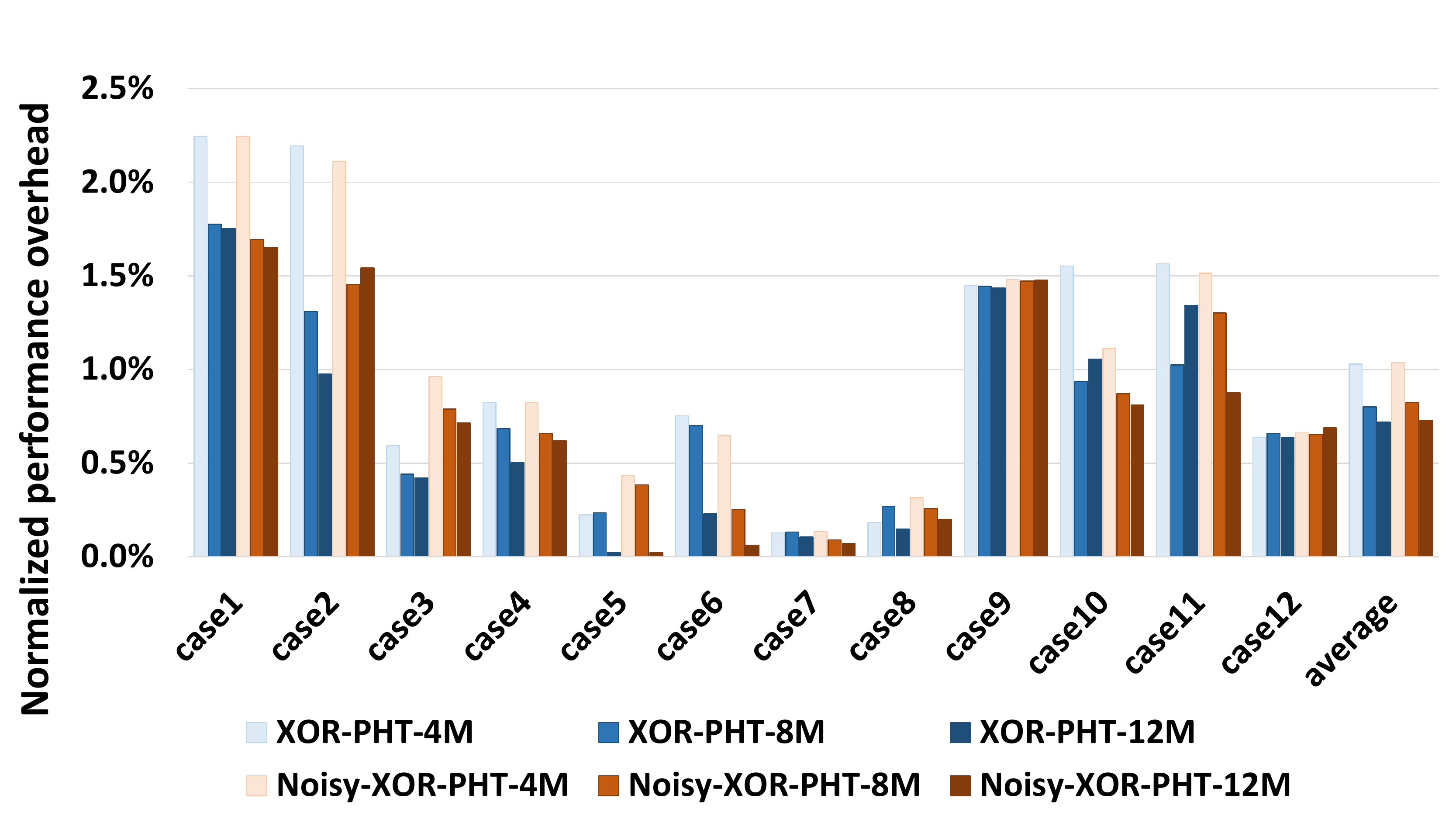}
\caption{Performance overhead of \emph{XOR-PHT} and \emph{Noisy-XOR-PHT}.}
\label{XOR_PHT}
\vspace{-1.4em}
\end{figure}

The performance overhead of \emph{XOR-PHT} is shown in Figure~\ref{XOR_PHT}, where
we see that the average overhead is less than 1.1\%. The performance loss decreases
gradually with the increase of the context switch interval. Following a key change,
the same thread would now face essentially randomized PHT content.
However, considering the 2-bit counter takes a short amount of time to warm up, 
the average performance overhead is relatively insignificant.
For example, the static conditional branch instruction ratios of case1  (gcc+calculix) are
relatively high at 12.1\% and 8.1\% respectively and
their accuracy of PHT prediction are 90.1\% and 94.0\% respectively, which
explains that the performance loss of the combination is the highest of our
twelve cases. Another example is case7 of single-threaded execution. The proportions of
static conditional branch instructions are just 4.8\% and 7.6\% for gromacs and GemsFDTD
respectively, and the accuracy of PHT prediction is 88.9\% for gromacs, and thus
every time the context switches, gromacs will scratch the results of GemsFDTD
training, so the \emph{XOR-PHT} mechanism has little impact on the combination.

\begin{figure}[ht]
 \vspace{-0.7em}
\centering
  \includegraphics[width=0.48\textwidth]{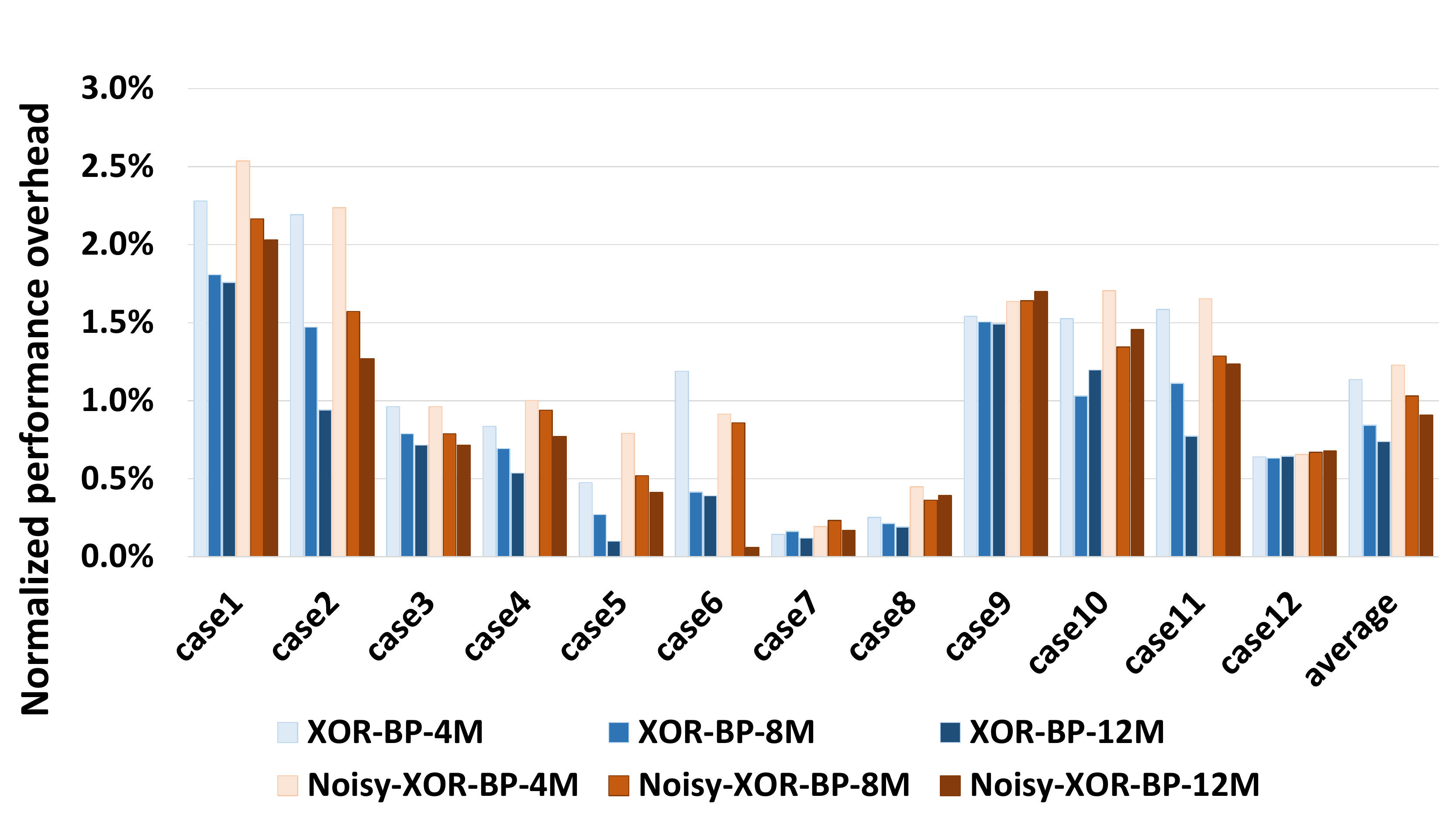}
\caption{Performance overhead of \emph{XOR-BP} and \emph{Noisy-XOR-BP}.}
\label{XOR_BP}
\vspace{-0.8em}
\end{figure}

When combining the protection on BTB and the direction predictor, the performance
impact is largely additive.
The performance impact of the resulting configuration \emph{Noisy-XOR-BP} is 
shown in Figure~\ref{XOR_BP}.
Overall, the average performance loss is less than 1.3\%. 
The largest performance loss  is about 2.5\% (case1), largely due to cost in 
PHT isolation seen earlier.

Overall, the additional impact of the XOR de/encoding mechanism is very small. And the main performance impact comes from PHT isolation.
It should be noted that this low performance overhead is similar to the results shown in Figure~\ref{flush_single}, and that the flush-based approach
is not expensive for single-threaded core.

\begin{scriptsize}
\begin{table}[!pt]
  \caption{The number of privilege switches per million cycles.}
  \label{Context_Priveledge_percentage}
\setlength{\abovecaptionskip}{0.1cm}
\setlength{\belowcaptionskip}{-0.25cm}
  \centering
  \setlength{\tabcolsep}{8pt}
  \small
  \begin{tabular}{|c|c|c|c|}
    \hline
    \multicolumn{1}{|c}{\textbf{Test}} & \multicolumn{1}{|c}{\textbf{Number of}} & \multicolumn{1}{|c}{\textbf{Test}} & \multicolumn{1}{|c|}{\textbf{Number of}}\\
    \multicolumn{1}{|c}{\textbf{Number}} & \multicolumn{1}{|c}{\textbf{privilege switches}} & \multicolumn{1}{|c}{\textbf{Number}} & \multicolumn{1}{|c|}{\textbf{privilege switches}}\\
    \hline
    \hline
        \multicolumn{1}{|c|}{case1} & 4.9 & {case7} & 1.7 \\ 
    \hline
    \multicolumn{1}{|c|}{case2} & 7.0 & {case8} & 2.0 \\
    \hline
    \multicolumn{1}{|c|}{case3} & 1.9 & {case9} & 1.8 \\
    \hline
    \multicolumn{1}{|c|}{case4} & 2.0 & {case10} & 2.7 \\
    \hline
    \multicolumn{1}{|c|}{case5} & 1.7 & {case11} & 3.5 \\
    \hline
    \multicolumn{1}{|c|}{case6} & 1.6 & {case12} & 1.9 \\
    \hline
  \end{tabular}
\vspace{-1.3em}
\end{table}
\end{scriptsize}

\subsubsection{Breakdown of two types of switches}

Take \emph{Noisy-XOR-BP}-12M as an example,
Table~\ref{Context_Priveledge_percentage} shows the number of privilege changes per million cycles.
We found that the number of privilege changes
is much larger than the number of context switches (0.08 per million cycles). Similar phenomena
can been observed in other cases. It is not difficult to understand that in
a time slice, the execution of the program may incur multiple privilege
changes due to system calls, exception handling etc. This
indicates that the main factor in determining the impact of isolation may
be the characteristics of the program itself, not the timer interruption
frequency setting. This also explains one observation in
Figure~\ref{XOR_BP}: As the timer switch frequency changes, there is no
significant fluctuation in the performance loss of \emph{Noisy-XOR-BP} for most test
cases.

\begin{figure*}[ht]
\centering
\hspace{1in}
\includegraphics[width=0.98\textwidth]{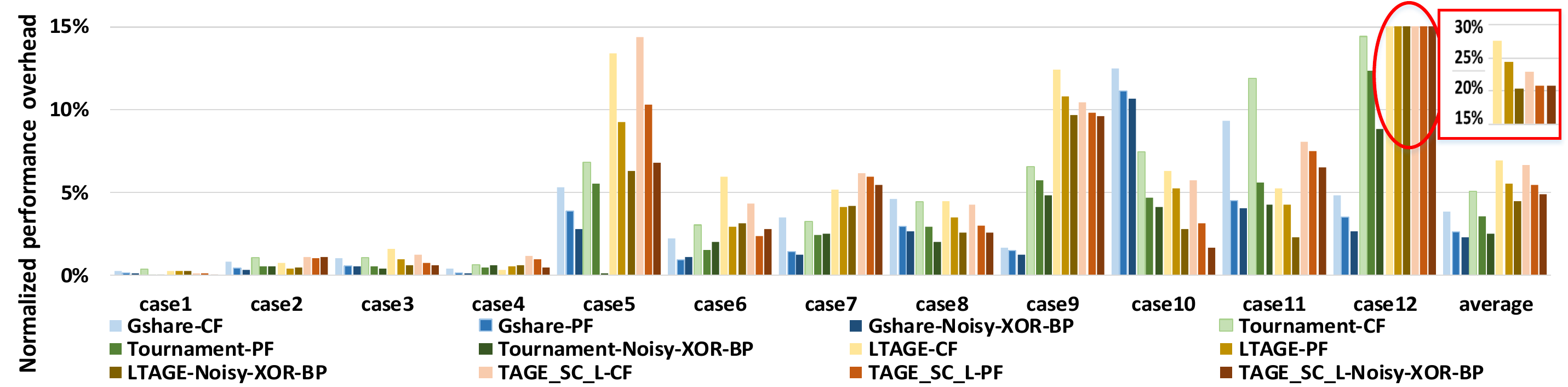}
\caption{Performance cost of three isolation mechanisms on four 
different predictors an SMT core. CF and PF refer to complete and precise
flush, respectively.}
\label{SMT_XOR_BP} %% label for entire figure
\end{figure*}

\subsection{Simulation-based evaluation on SMT cores}

Figure~\ref{SMT_XOR_BP} shows performance impacts of
three defence mechanisms (Complete flush, Precise flush, and Noisy-XOR-BP)
on four different branch predictors (Gshare, Tournament, LTAGE, and TAGE-SC-L,
with a measured baseline MPKI of 8.45, 5.17, 4.10, and 3.99 respectively).
Each bar is showing the performance degradation compared to the same predictor
without any protection. Three observations can be made.

\begin{enumerate}

\item There is a non-trivial range of performance impacts. In some cases, the
combination of application's sensitivity to branch predictor behavior and the
frequency of key changes can result in more than 20\% performance degradation.
But on an average, the performance cost for protection, at a few percent, is
quite reasonable.

\item Although there are exceptions, in general, Noisy-XOR-BP incurs lower
performance impact than both flush mechanisms. Compared to complete flush,
performance loss due to Noisy-XOR-BP is 26 to 37\% lower. Keep in mind
that the flush mechanisms are more costly to implement, and provides less
protection against attacks than our proposal.

\item Systems with a more accurate predictor tends to show more performance
impact due to protection. But on average, the increase is not dramatic: It
goes from 2.3\% for the least accurate predictor to 4.9\% to the
most accurate. 
\end{enumerate}

\subsection{Hardware Cost Estimation}

The \emph{XOR-BP} and \emph{Noisy-XOR-BTB} are implemented at Register-Transfer Level (RTL) on RISC-V processor. Based on TSMC 28nm technology, we use Synopsys ASIC design flow and synthesis tools to assess the timing and area cost. Table ~\ref{hardware_evaluation} shows the timing and area overhead of \emph{Noisy-XOR-BP} with different configurations. And the numbers are compared with original BTB and PHT. It can be seen that \emph{Noisy-XOR-BP} has minor area and timing cost. For example, in case of 2-way BTB with 256 entries each way, the timing cost of \emph{Noisy-XOR-BTB} is increased by 0.94\% and area cost is increased by 0.15\% (Synthesized with TT corner using Design Compiler). 

\vspace{-0.5em}
\begin{scriptsize}
\begin{table}[htb]
  \caption{Area and timing evaluation. }
  \label{hardware_evaluation}
\setlength{\abovecaptionskip}{-0.3em}
\setlength{\belowcaptionskip}{-0.25cm}
  \centering
%   \begin{group}
  \setlength{\tabcolsep}{12pt}
%   \linespread{1.3}
  \small
  \begin{tabular}{|c|c|c|c|}
    \hline
    \multicolumn{4}{|c|}{BTB (2w128 means 2-way with 128 entries in each way)} \\
    \hline
    & 2w128 & 2w256 & 2w512 \\
    \hline
    Timing & 0.70\% & 0.94\% & 1.46\% \\
    \hline
    Area & 0.24\% & 0.15\% & 0.13\% \\
    \hline
    \hline
    \multicolumn{4}{|c|}{PHT (TAGE Predictor)} \\
    \hline
    & 1024 entries & 2048 entries & 4096 entries\\
    & per table & per table & per table\\
    \hline
    Timing & 2.10\% & 1.98\% & 2.01\% \\
    \hline
    Area & 0.11\% & 0.09\% & 0.03\% \\
    \hline    
  \end{tabular}
%   \end{group}
\vspace{-1.3em}
\end{table}
\end{scriptsize}

\section{Related Work}
\label{related}
\subsection{Comparison with cache side channel attacks}

There are two type of cache side channels: contention based channels and 
reuse based channels~\cite{liu2014random}. For contention based channels, there is a meaningful map between addresses and entries in the cache or
table to allow the attacker to achieve manipulation. Randomizing the mapping,
therefore, is one way to increase the barrier of attack. RPCache~\cite{wang2007new},
CEASER~\cite{qureshi2018ceaser}, and ScatterCache~\cite{wernerscattercache}
all use some form of randomization. 
For a reuse based channel, a
typical attack is the Flush+Reload~\cite{yarom2014flush}. Through flushing
specific lines, the attacker can force lengthy memory accesses on certain
addresses shared between the attacker and the victim. Countermeasures to such
attacks include limiting the use of flush or de-correlate caching from demand
access via Random Fill Cache~\cite{liu2014random}.

Unlike Cache, branch prediction tables need both index and content encoding. The content encoding only affects the accuracy of the branch predictor, not
the correctness of the programs.

\subsection{Countermeasures for branch predictor attacks}

Several countermeasures have been proposed. First, we can reduce information
leakage through the side channel. For instance, branches that conclude
sensitive information can be transformed into safe instructions that do not
leave a mark in the branch predictors~\cite{agosta2007countermeasures}.
Limiting performance counter usage can reduce the information obtained by
the attacker~\cite{guide2011intel}. InvisiSpec~\cite{yan2018invisispec},
SafeSpec~\cite{khasawneh2019safespec}, MuonTrap~\cite{sam2020muontrap}, Conditional
Speculation~\cite{li2019conditional}, Efficient Invisible Speculative Execution~\cite{sakalis2019efficient}, CleanupSpec~\cite{saileshwar2019cleanupspec}, STT~\cite{yu2019speculative}, ConTExT~\cite{schwarz2020context}, Speculative Data-Oblivious Execution~\cite{jiyong2020speculative} and SpecShield~\cite{barber2019specshield} prevent speculative computation
from generating visible microarchitectural state. Alternatively, Retpoline~\cite{turner2018retpoline},
FENCE~\cite{Intel2018Mitigations,ARM2018Mitigations,AMD2018Mitigations}, and
Context-Sensitive Fencing~\cite{taram2019context} mitigate the chance that
untrusted code is speculatively executed.

Second, the predictor table can be flushed to contain randomized new
results. Performing this by software during context switch can bring
non-trivial overhead~\cite{evtyushkin2016understanding}. Such expensive operations can be limited to only the sensitive processes~\cite{hu1992lattice}. The impact on performance and
prediction accuracy of flushing predictor table in hardware has been
studied~\cite{evers1996using, pasricha2003improving}. Not surprisingly,
the longer the context switch interval, the smaller the impacts. These
observations are consistent with ours.

Third, using more dedicated hardware is a general approach to isolate states
from different processes. For example, sensitive applications in SGX can be
allocated with their own branch predictor tables~\cite{evtyushkin2018branchscope}.  Earlier work on
performance improvement considered saving and restoring compressed branch
prediction information~\cite{dhodapkar2001saving} or providing thread-private
branch predictors on SMT processors~\cite{ramsay2003exploring}. BRB is a proposal
to retain partial predictor state in on-chip SRAM to swap in with
context~\cite{vougioukas2019brb}.

Finally, detecting (and subsequently freezing or killing)
malicious processes is a possible general defense
mechanism~\cite{evtyushkin2018branchscope}. While some specific methods have
been discussed~\cite{gianvecchio2007detecting}, accurately detecting malicious
processes remains a difficult, not-yet-practical approach.

% % that's all folks

\section{Conclusion}
\label{conclusion}

Branch prediction is crucial in modern high-performance microprocessors. But
the basic design that allows sharing of branch predictors between threads
leaves opportunities for malicious training and perception.

Instead of traditional approaches based on flushing or using separate hardware
resources, this paper explores isolation of the \emph{content} in hardware
tables via (1) content and (2) index encoding. In content encoding, we use
hardware-based thread-private random numbers to encode both direction and
destination histories. With index encoding, we use a another thread-private
random number as part of the input to compute the index of the tables,
disrupting the correspondence between the branch instruction address and
its table entry. The combination of the actions achieves excellent logical
isolation of branch histories between threads and privilege levels while
adding little in terms of circuit area or delay of critical circuit path.

Our evaluation using an FPGA-based RISC-V processor prototype and additional
auxiliary simulations demonstrate that the proposed mechanism adds less than
5\% slowdown on average. Compared to flush-based protection mechanisms that
have less protection in SMT cores and can incur far more circuit area, our
design reduces performance penalty by about 20-50\% depending on the branch
predictor.

% that's all folks

%%%%%%% -- PAPER CONTENT ENDS -- %%%%%%%%

%%%%%%%%% -- BIB STYLE AND FILE -- %%%%%%%%
\bibliographystyle{IEEEtranS}
\bibliography{refs}
%%%%%%%%%%%%%%%%%%%%%%%%%%%%%%%%%%%%

\end{document}